\documentclass[conference,9pt]{IEEEtran}
\IEEEoverridecommandlockouts
\usepackage{amsmath,amssymb,amsfonts}
\usepackage[ruled, vlined, linesnumbered]{algorithm2e}
\usepackage[version=4]{mhchem}
\usepackage{graphicx}
\usepackage{textcomp}
\usepackage{xcolor}
\usepackage{booktabs}

\usepackage{tikz}
\usetikzlibrary{quantikz2,shapes.geometric, arrows, positioning, fit, calc}

\usepackage{subfigure}
\usepackage{hyperref}
\usepackage{makecell}
\usepackage{adjustbox}
\usepackage{cleveref}
\usepackage{multirow}
\usepackage{soul}

\Crefname{figure}{Fig.}{Figs.}
\Crefname{table}{Tab.}{Tabs.}
\crefname{equation}{Eq.}{Eqs.}

\newcommand{\dquote}[1]{``#1''}
\newcommand{\code}{\texttt}

\newcommand{\phoenix}{\textsc{Phoenix}}
\newcommand{\qiskit}{\textsc{Qiskit}}
\newcommand{\tket}{\textsc{TKet}}
\newcommand{\tetris}{\textsc{Tetris}}
\newcommand{\paulihedral}{\textsc{Paulihedral}}
\newcommand{\pauliopt}{\textsc{PauliOpt}}
\newcommand{\pcoast}{\textsc{PCOAST}}
\newcommand{\twoqan}{\textsc{2QAN}}

\newcommand{\CHtwo}{CH$_2$}
\newcommand{\HtwoO}{H$_2$O}
\newcommand{\LiH}{LiH}
\newcommand{\NH}{NH}

\newcommand{\SWAP}{\mathrm{SWAP}}
\newcommand{\CNOT}{\mathrm{CNOT}}
\newcommand{\SUfour}{\mathrm{SU}(4)}

\newcommand{\totalWeight}{w_\mathrm{tot.}}

\newcommand{\eRrightPre}{e_r}
\newcommand{\eLeftPost}{e_l'}

\SetAlFnt{\small}
\SetCommentSty{textit}

\hypersetup{hidelinks}

\def\BibTeX{{\rm B\kern-.05em{\sc i\kern-.025em b}\kern-.08em
    T\kern-.1667em\lower.7ex\hbox{E}\kern-.125emX}}

\begin{document}


\title{\phoenix: Pauli-Based High-Level Optimization Engine for Instruction Execution on NISQ Devices}


 \author{
     \IEEEauthorblockN{Zhaohui Yang\IEEEauthorrefmark{1}, Dawei Ding\IEEEauthorrefmark{2}, Chenghong Zhu\IEEEauthorrefmark{3}, Jianxin Chen\IEEEauthorrefmark{4}\thanks{\textsuperscript{\S}Corresponding author. Email: \href{mailto:chenjianxin@tsinghua.edu.cn}{chenjianxin@tsinghua.edu.cn}.}, Yuan Xie\IEEEauthorrefmark{1}}
    
     \IEEEauthorblockA{\textit{\IEEEauthorrefmark{1}Department of Electronic and Computer Engineering, The Hong Kong University of Science and Technology, Hong Kong}}

     \IEEEauthorblockA{\textit{\IEEEauthorrefmark{2}Yau Mathematical Sciences Center, Tsinghua University, Beijing 100084, China}}

     \IEEEauthorblockA{\textit{\IEEEauthorrefmark{3}The Hong Kong University of Science and Technology (Guangzhou), Guangdong 511453, China}}

     \IEEEauthorblockA{\textit{\IEEEauthorrefmark{4}Department of Computer Science and Technology, Tsinghua University, Beijing 100084, China}}


}

\maketitle

\begin{abstract}
    Variational quantum algorithms (VQA) based on Hamiltonian simulation represent a specialized class of quantum programs well-suited for near-term quantum computing applications due to its modest resource requirements in terms of qubits and circuit depth. Unlike the conventional single-qubit (1Q) and two-qubit (2Q) gate sequence representation, Hamiltonian simulation programs are essentially composed of disciplined subroutines known as Pauli exponentiations (Pauli strings with coefficients) that are variably arranged. To capitalize on these distinct program features, this study introduces \phoenix, a highly effective compilation framework that primarily operates at the high-level Pauli-based intermediate representation (IR) for generic Hamiltonian simulation programs. \phoenix\ exploits global program optimization opportunities to the greatest extent, compared to existing SOTA methods despite some of them also utilizing similar IRs. \phoenix\ employs the binary symplectic form (BSF) to formally describe Pauli strings and reformulates IR synthesis as reducing the column weights of BSF by appropriate Clifford transformations. It comes with a heuristic BSF simplification algorithm that searches for the most appropriate 2Q Clifford operators in sequence to maximally simplify the BSF at each step, until the BSF can be directly synthesized by basic 1Q and 2Q gates. \phoenix\ further performs a global ordering strategy in a Tetris-like fashion for these simplified IR groups, carefully balancing optimization opportunities for gate cancellation, minimizing circuit depth, and managing qubit routing overhead. Experimental results demonstrate that \phoenix\ outperforms SOTA VQA compilers across diverse program categories, backend ISAs, and hardware topologies.
    
\end{abstract}




\section{Introduction}

    Quantum computing offers the potential to revolutionize various fields, driving decades of efforts to develop the required physical hardware. For instance, quantum algorithms can achieve exponential speedups in tasks such as integer factorization~\cite{shor1994algorithms}, solving linear equations~\cite{harrow2009quantum}, and quantum system simulation~\cite{lloyd1996universal}. In the noisy intermediate-scale quantum (NISQ) era where we have access to dozens or hundreds of qubits susceptible to noise (e.g., qubit decoherence, gate imperfections)~\cite{preskill2018quantum}, variational quantum algorithms (VQA) is a leading class of algorithms proposed to achieve the quantum advantage (e.g., VQE for chemistry and condensed-matter simulation~\cite{peruzzo2014variational}, QAOA for combinatorial optimization~\cite{farhi2014quantum}) due to its modest resource requirements in terms of qubit number and circuit depth as well as its noise-resilience.~\cite{cerezo2021variational}.


    The construction of a VQA ansatz circuit is to simulate (approximate) a desired unitary evolution under the system Hamiltonian $ H $ and the evolution duration $ t $, through Trotterizing~\cite{cerezo2021variational} the evolution $ U(t) $ given $ H $ represented by a linear combination of Pauli strings:
    \begin{align}
        U(t) = e^{-iHt} &\simeq  \left( S_{k}\left(\tau\right) \right)^r,\, \tau = \frac{t}{r},\\
        H = \sum\nolimits_{j=1}^L h_j P_j =& \sum\nolimits_{j=1}^L h_j \sigma_0^{(j)} \otimes \cdots \otimes \sigma_{n-1}^{(j)},
    \end{align}
    where $ k $ and $ r $ indicate the Trotter order and time step, respectively. Finer-grained Trotterization results in lower approximation errors. For example, the 1st-order and 2nd-order Trotterization are given by
    \begin{align*}
        S_1 = \prod\nolimits_{j=1}^{L} e^{ -i h_j \tau P_j},\, S_2 = \prod\nolimits_{j=1}^{L} e^{ -i h_j \frac{\tau}{2} P_j}\prod\nolimits_{j=L}^1 e^{ -i h_j \frac{\tau}{2} P_j},
    \end{align*}
    respectively.
Following Trotterization, each term $S_k$ is expressed as a product of individual Pauli exponentiations, which can be easily synthesized using basic 1Q and 2Q gates. The arrangement of these Pauli exponentiations within each Trotter step can be freely chosen without affecting the upper bound of Trotterization (approximation) error~\cite{cerezo2021variational}, which is usually not comparable to physical noise. However, different arrangements offer varying opportunities for optimizing quantum circuits, thereby dominating the accuracy of Hamiltonian simulations on noisy hardware. Consequently, the primary challenge of compiling VQA ansatz circuits lies in synthesizing these Pauli exponentiations from a global perspective, dubbed \emph{Pauli-based intermediate representations (IR)} throughout this paper.

    \begin{figure}[tbp]
        \centering   \includegraphics[width=\columnwidth]{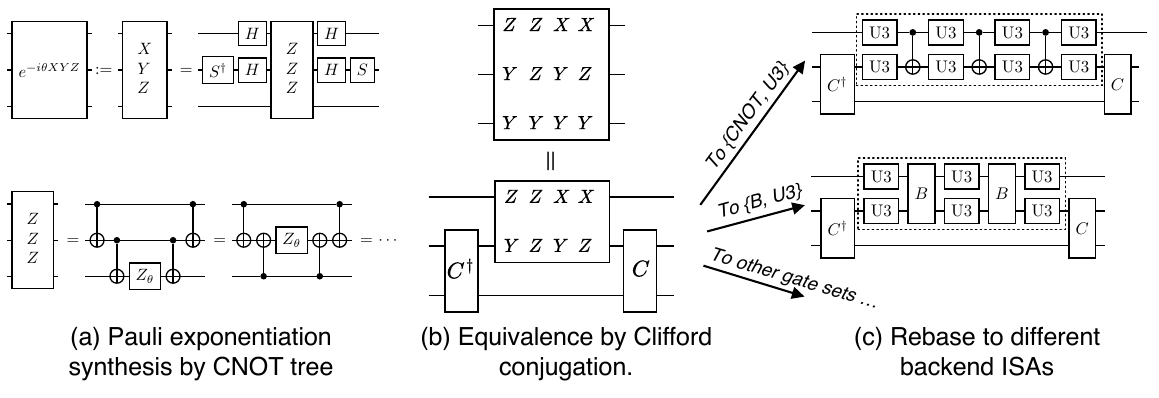}
        \caption{Conventional Pauli exponentiation synthesis v.s. Simultaneous simplification by Clifford conjugation. (a) Pauli exponentiations are synthesized by gate set $ \{ H, S, S^\dagger, Z(\theta), \CNOT \} $, through variable $ \CNOT $-tree unrolling schemes. (b) Multiple 3Q Pauli exponentiations can be simultaneously simplified into 2Q Pauli exponentiations, through a 2Q Clifford conjugation, where $ C = (H\otimes S)\, \CNOT\, (H\otimes S^\dagger) $ in this example. (c) The simplified Pauli exponentiations can be rebased to versatile quantum gate sets, such as $ \CNOT $-based and $ \mathrm{B} $-based ISAs, with significantly reduced 2Q gate count.}
        \label{fig:motivation}
        
    \end{figure}

    Conventionally, a Pauli exponentiation is synthesized into a 1Q rotation $ Z(\theta) $ sandwiched by a pair of symmetric $ \CNOT $ trees, subsequently conjugated by some $ H $ and $ S $ gates, as exemplified in \Cref{fig:motivation} (a). Existing state-of-the-art (SOTA) compilers primarily exploit gate cancellation opportunities exposed by the already synthesized subcircuits, whether by means of the abstract ZX diagram~\cite{cowtan2019phase,van2023towards,paykin2023pcoast} representation or the variants of $ \CNOT $ trees~\cite{li2022paulihedral,jin2024tetris}. Despite utilizing the Pauli-based IR, the optimization process they formulate is limited to subcircuits and local IR patterns. Furthermore, it assumes the conventional $ \CNOT $-based quantum instruction set architecture (ISA). We instead present that a set of Pauli exponentiations \emph{can be simultaneously simplified} through appropriate Clifford transformations. For example, \Cref{fig:motivation} (b) shows that the list of weight-3 Pauli strings $ [ZYY;\, ZZY;\, XYY;\, XZY] $ can be simplified into a weight-2 Pauli string list through the conjugation of a 2Q Clifford operator $ C = (H\otimes S)\, \CNOT\, (H\otimes S^\dagger) $. This approach unlocks greater optimization opportunities \emph{entirely at the level of Pauli-based IR} while remaining agnostic to the quantum ISA being employed.

    In this work, we propose \phoenix\ \emph{(Pauli-based High-level Optimization ENgine for Instruction eXecution)}---a highly effective compilation framework for generic Hamiltonian simulation programs on near-term quantum devices. \phoenix\ follows the 
    \begin{align*}
        \text{\dquote{IR grouping $\to$ group-wise simplification $\to$ IR group ordering}}
    \end{align*}
    pipeline to compile real-world VQA programs into basic quantum gates. It is ISA-independent, routing-aware, and tailored to generic VQA programs such as molecular simulation involving heterogeneous-weight Pauli strings and QAOA including only weight-2 Pauli strings. Differing from existing SOTA methods, \phoenix\ utilizes an alternative formal description for Pauli-based IR and integrates heuristic optimization strategies to achieve global optimization primarily at the high-level semantic layer. We evaluate \phoenix\ across diverse VQA programs, backend ISAs, and hardware topologies, demonstrating its superior performance over existing SOTA compilers. Our main contributions are as follows:
    \begin{enumerate}
        \item We utilize the \emph{binary symplectic form} (BSF) as the formal description of Pauli-based IR and reformulate the IR synthesis process as simultaneous simplification on BSF by appropriate Clifford conjugations that simultaneously reduce the BSF's column weights, enabling global optimization to the largest extent while operating in the high-level IR.
        \item We propose a heuristic \textit{BSF simplification} algorithm as the core optimization pass in \phoenix. It iteratively searches for the most appropriate 2Q Clifford operators to quickly lower the weight of the BSF, until the BSF can be directly synthesized by basic 1Q and 2Q gates.
        \item We quantify the circuit depth overhead induced by assembling (ordering) simplified IR groups. A uniform cost function is devised to comprehensively account for circuit depth overhead, gate cancellation, and qubit routing overhead. Under the guidance of that, \phoenix\ integrates a \emph{Tetris-like IR group ordering} procedure to reliably achieve a good ordering scheme.
    \end{enumerate}

    Overall, \phoenix\ outperforms the best-known compilers (e.g., \tket~\cite{sivarajah2020t}, \paulihedral~\cite{li2022paulihedral}, \tetris~\cite{jin2024tetris}), achieving significant reductions in gate count and circuit depth. For example, for logical-level compilation, \phoenix\ results in 80.47\% reduction in $ \CNOT $ gate count and 82.7\% reduction in 2Q circuit depth on average, compared to the original logical circuits. For hardware-aware compilation with heavy-hex topology, \phoenix\ reduces by 36.17\% (22.62\%) in $ \CNOT $ gate count and 43.85\% (28.12\%) in 2Q circuit depth on average, compared to \paulihedral\ (\tetris). The reduction effect in 2Q gate count and circuit depth becomes even more impressive when targeting the newly introduced $ \SUfour $ ISA (the representative continuous ISA containing all 2Q gates)~\cite{chen2024one}.




\section{Related Works}

    \textbf{Based on ZX diagram representations of Pauli gadgets.} Compilers such as \tket~\cite{sivarajah2020t,cowtan2019phase}, \pauliopt~\cite{van2023towards}, and \pcoast~\cite{paykin2023pcoast}\ resynthesize quantum circuits into its ZX diagram representation. ZX diagrams are used to visually represent and simplify the commutation relations and phase interactions between Pauli operators, facilitating the reduction of circuit depth and gate count through algebraic and graphical transformations. Although it can efficiently resynthesize Pauli gadgets, the commutation rules it leverages occur locally, and it is hard to operate in a hardware-aware manner.

    \textbf{Based on synthesis variants of Pauli-based IR.} Dedicated compilers like \paulihedral\cite{li2022paulihedral} and \tetris~\cite{jin2024tetris} leverage Pauli-based IR to identify gate cancellation opportunities between nearest-neighbor IRs, exposed by the variants of their synthesis schemes based on $\CNOT$-tree unrolling. They proposed sophisticated co-optimization techniques to minimize $ \CNOT $ gate for both logical-level synthesis and $\SWAP$-based routing achieve good performance especially on limited-topology NISQ devices. However, their optimization scope is confined to a finite set of local subcircuit patterns, and they rely solely on the $ \CNOT $-based ISA.


    













\section{BSF and Clifford Formalism}\label{sec:bsf-clifford}

    \begin{figure}[tbp]
        \centering
        \includegraphics[width=\columnwidth,trim={0 0.25cm 0 0},clip]{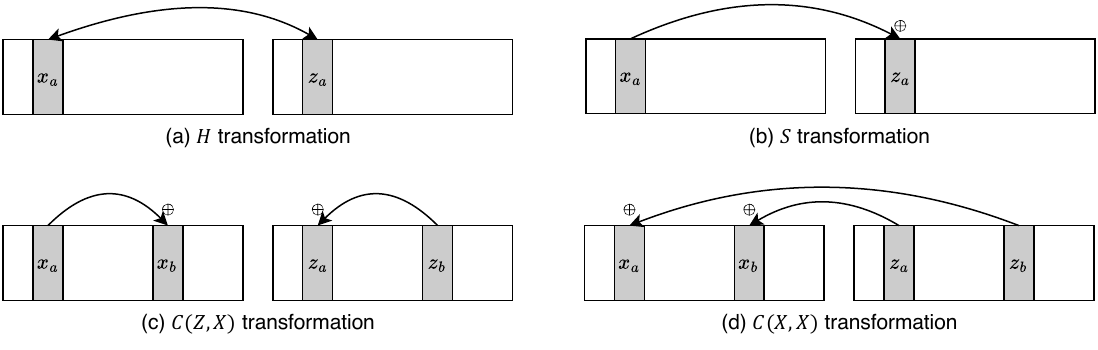}
        
        \caption{Examples of Clifford transformations on BSF. Columns $a$ and $b$, corresponding to qubit $a$ and $b$, for $X$ and $Z$ blocks are indicated by $x_a$, $x_b$, $z_a$, $z_b$, respectively. (a) $H$ gate acting on qubit $a$ will exchanges $x_a$ and $z_a$. (b) $S$ gate acting on qubit $a$ results in $z_a \gets z_a \oplus x_a$. (c) $C(Z,X)_{a,b}$, i.e., the $\CNOT$ gate, results in $x_b \gets x_b \oplus x_a$ and $z_a \gets z_a \oplus z_b$. (d) $C(X,X)$ transformation is equivalent with successively applying $H$ on $a$, $C(Z,X)_{a,b}$, and $H$ on $a$, resulting in $x_a \gets x_a \oplus z_b$ and $x_b \gets x_b \oplus z_a$.}
        \label{fig:clifford}
    \end{figure}


    We establish our synthesis scheme by representing Pauli strings in the binary symplectic form (BSF). In the BSF, each $n$-qubit Pauli string is represented as a row in a tableau, with columns divided into two sections: $[X\,|\, Z]$. Here, $[X_{i,j} \,|\, Z_{i,j}]$ denotes the $j$-th component of the $i$-th Pauli operator. The encoding maps $X$ to $[1 \,|\, 0]$, $Z$ to $[0 \,|\, 1]$, $Y$ to $[1 \,|\, 1]$, and $I$ to $[0 \,|\, 0]$. For instance, the Pauli string simplification shown in \Cref{fig:motivation} (b) follows the formulation
    \begin{align*}
        \left[
            \begin{array}{ccc|ccc}
                0 & 1 & 1 & 1 & 1 & 1\\
                0 & 0 & 1 & 1 & 1 & 1\\
                1 & 1 & 1 & 0 & 1 & 1\\
                1 & 0 & 1 & 0 & 1 & 1\\
            \end{array}
        \right]
        \xrightarrow{C(X,Y)_{1,2}}
        \left[
            \begin{array}{ccc|ccc}
                0 & 1 & 0 & 1 & 1 & 0\\
                0 & 0 & 0 & 1 & 1 & 0\\
                1 & 1 & 0 & 0 & 1 & 0\\
                1 & 0 & 0 & 0 & 1 & 0\\
            \end{array}
        \right],
    \end{align*}
    where $ C(X,Y)_{1,2} $ is the Clifford gate in \Cref{fig:motivation} (b) while written in the \emph{universal controlled gate} representation
    \begin{align*}
        C(\sigma_0, \sigma_1) = \frac{1}{2}((I + \sigma_0)\otimes I + (I - \sigma_0)\otimes \sigma_1),\, \sigma_j\in \left\{X,Y,Z\right\}.
    \end{align*}
    Any universal controlled gate is equivalent to $ \CNOT $ up to local $ H $ and $ S $ conjugations. For example, $ C(Z,X) = \CNOT$, and $ C(X,Y)  =  (H\otimes S)\, \CNOT\, (H\otimes S^\dagger) $.

    By definition, any operator $ C $ from the Clifford group is unitary and has the property that $ C P C^\dagger $ is also a Pauli operator for any Pauli operator $ P $. The update rules of some Clifford operators in terms of the BSF are demonstrated in \Cref{fig:clifford}, in which only 2Q Clifford operators (Clifford2Q) potentially have nontrivial effects of reducing the Pauli strings' weights. For any $ C(\sigma_0, \sigma_1) $, its tableau update rule is a combinatorial sequence of $ H $, $ S $ and $ C(Z, X) $ (or $ C(Z, Z) $) transformations. For example, $C(X,Y)_{a,b}$ exhibits the rule
    \begin{align}
    [x_a,\, x_b\, |\, z_a,\, z_b] \rightarrow [x_a\oplus x_b\oplus z_b,\, z_a\oplus z_b\, |\, z_a,\, z_a\oplus z_b].
    \end{align}
        
    Therefore, the basic approach of our synthesis scheme is to find an appropriate sequence of Clifford2Q operators to simplify the BSF\footnote{The BSF that needs to simplify does not contain any weight-1 Pauli string. } until it can be directly synthesized by basic 1Q and 2Q gates, i.e., the quantity \dquote{total weight} 
    \begin{align}
        \totalWeight \coloneqq \lVert  \lor_i (r_x^{(i)}\lor r_z^{(i)})  \rVert
    \end{align}
    is no more than 2, where $r_{x/z}^{(i)}$ is the $i$-th row of $X$ or $Z$ block of the BSF, and the norm is the sum of all binary entries. The key lies in how to find the Clifford2Q that can reduce the weights of as many Pauli strings as possible within the BSF.


\section{Our Proposal: \phoenix}

\subsection{Overview}

    In the proposed framework \phoenix, Pauli-based IRs (exponentiations) are first grouped according to the same set of qubit indices non-trivially acted on. Then, the \emph{BSF simplification} algorithm is applied to the BSF of each IR group, generating the simplified subcircuit composed of $ \CNOT $-equivalent Clifford2Q operators and Pauli exponentiations with weights no more than 2. \phoenix\ further selectively assemble (order) these unarranged simplified IR groups in a Tetris-like style that minimizes a uniform subcircuit assembling cost function that incorporates gate cancellation, circuit depth increase, and qubit routing overhead. By defining appropriate metric functions and heuristic algorithms, \phoenix\ achieves superior optimization (less 2Q gate count and circuit depth) compared to SOTA methods, efficiently handling compilation for large-scale VQA programs. 

\subsection{BSF simplification for each IR group}

    \begin{algorithm}[tbp]
    \SetAlgoLined
    \caption{Pauli Strings Simplification in BSF}
    \label{algo:simplification}
    \SetKwInOut{Input}{Input}
    \SetKwInOut{Output}{Output}
    \SetKwBlock{Assumption}{Assumption}{}

    \Input{Pauli strings list \textit{pls}}
    \Output{Reconfigured circuit components list \textit{cfg}}

    \BlankLine
    \textit{cfg} $ \gets \emptyset $;\quad
    \textit{bsf} $ \gets \textsc{BSF}$(\textit{pls});\quad
    \textit{cliffs\_with\_locals} $\gets \emptyset$\; 
    \While{bsf.\textsc{totalWeight()} $>$ 2}{
        \textit{local\_bsf} $\gets$ \textit{bsf}.\textsc{popLocalPaulis}()\; 

        $ C \gets \emptyset$ \tcp*{Clifford2Q candidates}
        $ B \gets \emptyset$ \tcp*{Each element of $B$ results from applying each Clifford2Q candidate on \textit{bsf}}
        \textit{costs} $\gets \emptyset$ \tcp*{Cost functions calculated on each element of $B$}
        \For{cg \textbf{in} \textsc{CLIFFORD\_2Q\_SET}}{
            \For{i, j \textbf{in} $ \textsc{combinations}(\textsc{range}(n), 2) $}{
                \textit{cliff} $\gets$ \textit{cg}.\textsc{on}$ (i, j) $ \tcp*{qubits acted on}
                \textit{bsf}$'$ $\gets$ \textit{bsf}.\textsc{applyClifford2Q}(\textit{cliff})\;
                \textit{cost} $\gets$ \textsc{calculateBSFCost}(\textit{bsf}$'$)\;
                $ C.\textsc{append}$(\textit{cliff})\;
                $ B.\textsc{append}$(\textit{bsf}$'$)\;
                \textit{costs}.\textsc{append}(\textit{cost})\;
            }
        }
        \textit{bsf} $ \gets \textsc{BSFWithMinCost} (B, \textit{costs}) $\;
        \textit{cliff} $ \gets \textsc{CliffordWithMinCost} (C, \textit{costs}) $\;
        \textit{cliffs\_with\_locals}.\textsc{append}((\textit{cliff}, \textit{local\_bsf}))\;
    }
    \BlankLine
    \textit{cfg}.\textsc{append}(\textit{bsf})\;
    \For{cliff, local\_bsf \textbf{in} cliffs\_with\_locals}{
        \tcp{Clifford2Q operators are added as conjugations, with local Pauli strings peeled before each epoch}
        \textit{cfg}.\textsc{prepend}(\textit{cliff})\;
        \textit{cfg}.\textsc{append}(\textit{local\_bsf})\;
        \textit{cfg}.\textsc{append}(\textit{cliff})\;
    }

\end{algorithm}

    To search for appropriate Clifford2Q operators in BSF simplification, it suffices to just focus on a set of Clifford2Q generators of the 2Q Clifford group. We choose the six universal controlled gates
    \begin{align}
        \left\{ C(X,X), C(Y,Y), C(Z,Z), C(X,Y), C(Y,Z), C(Z,X) \right\}\label{eq:clifford-generators}
    \end{align}
    that are independent of each other and span the entire 2Q Clifford group~\cite{grier2022classification}. It is a natural choice of sets of Clifford2Q generators, as each of them is Hermitian and equivalent to $ \CNOT $, and its corresponding BSF's tableau update rule can be defined as a combination of the update rules for $ H $, $ S $, and $ \CNOT $. 

    To determine which Clifford2Q operator to select from \Cref{eq:clifford-generators}, we define a heuristic cost function
    \begin{align}
        \mathrm{cost}_{\mathrm{bsf}} \coloneqq \, & \totalWeight * n_{\mathrm{n.l.}}^2 \notag + \sum\nolimits_{\langle i,j \rangle} \lVert r_x^{(i)} \lor r_z^{(i)} \lor r_x^{(j)} \lor r_z^{(j)} \rVert  \notag\\
        & \quad + \frac{1}{2} \sum\nolimits_{\langle i,j \rangle} (\lVert r_x^{(i)} \lor r_x^{(j)} \rVert + \lVert r_z^{(i)} \lor r_z^{(j)} \rVert) \label{eq:cost-bsf} 
    \end{align}
    to quantify the disparity of the current BSF from a desired BSF that requires no further simplification ($ \totalWeight \leq 2 $). $ \mathrm{cost}_{\mathrm{bsf}} $ is the combined weight overlap of both $ X $-part and $ Z $-part among each pair of Pauli strings of a BSF, with a bias considering the impact of the number of nonlocal (n.l., that is, weight larger than 1) Pauli strings. 
    
    \Cref{algo:simplification} illustrates the BSF simplification procedure in detail. It takes set of Pauli strings as input and outputs a configuration sequence \textit{cfg}, in which each component is either a Clifford2Q generator from \Cref{eq:clifford-generators} or a BSF with $\totalWeight$ at most 2. Specifically, at each Clifford2Q search epoch, Cref{algo:simplification} identifies the Clifford 2Q generator from \Cref{eq:clifford-generators} and qubit pair to act upon that leads to the greatest reduction in the cost function defined in \Cref{eq:cost-bsf}, and the BSF is updated accordingly. This greedy process continues iteratively until the BSF's $ \totalWeight $ is no more than 2. Before each search epoch, local Pauli strings are peeled from the BSF, as they represent 1Q Pauli rotations and do not induce synthesis overhead. The output \textit{cfg} represents a simplified IR group, while still in high-level semantics (Clifford2Q, 1Q Pauli-based IRs, and successive 2Q Pauli-based IRs), independent of any specific quantum ISA.

    In theory, simplifying a BSF is always achievable, for example, we can reduce an individual Pauli string to be weight-1, peel it off, and then iterate. The efficacy of \Cref{algo:simplification} lies in its simultaneous simplification mechanism guided by the cost function in \Cref{eq:cost-bsf}.

\subsection{Ordering IR groups in a Tetris-like style}

    \begin{figure}[tbp]
        \centering        
        \includegraphics[width=\columnwidth]{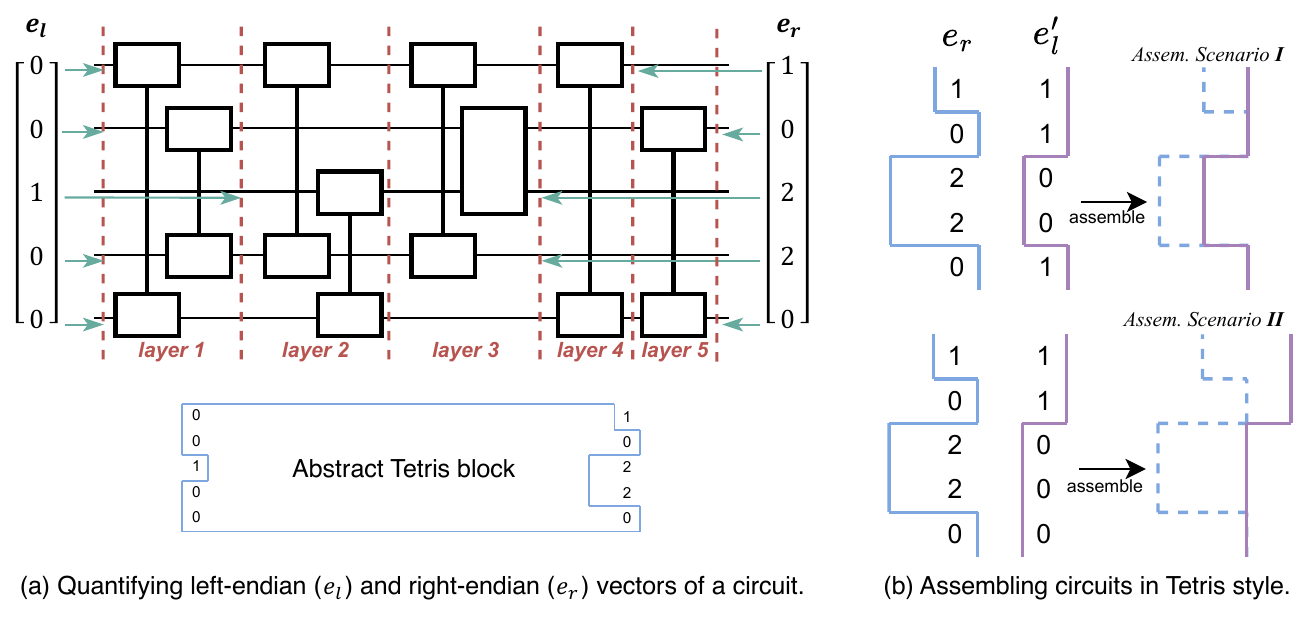}
        \caption{Tetris-like circuits assembling. (a) $ e_l $ and $ e_r $ examples. (b) Scenario I: $ \mathrm{cost}_\mathrm{depth} = \textsc{sum}(\eRrightPre + \eLeftPost) $; Scenario II: $ \mathrm{cost}_\mathrm{depth} = \textsc{sum}(\eRrightPre + \eLeftPost - 1)  $.}
        \label{fig:empty-end}
    \end{figure}

\begin{figure}[tbp]
        \centering
        \includegraphics[width=\columnwidth]{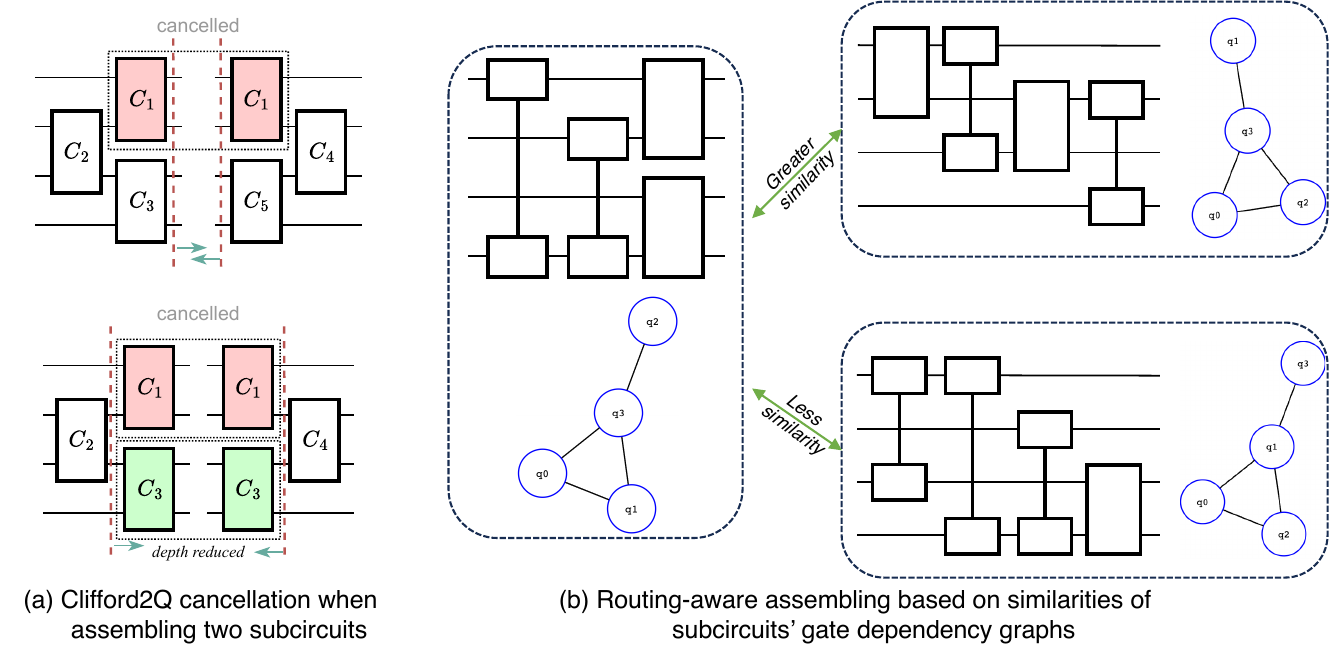}
        \caption{Gate cancellation opportunities and routing-aware assembling.  (a) Clifford2Q cancellation between the preceding and succeeding subcircuits may or may not induce circuit depth decrease.  (b) The subcircuit (upper right) whose qubit interaction graph is more similar to that of the already assembled subcircuit (left) is preferred over the other (lower right).}
        \label{fig:gate-cancel}
    \end{figure}

    With simplified IR groups compiled via BSF simplification, \phoenix\ further performs a \emph{Tetris-like IR group ordering} procedure that aims to minimize \textit{a uniform assembling cost metric}. This is unlike \cite{li2022paulihedral} and \cite{jin2024tetris} where the ordering is primarily determined by gate cancellation opportunities between IR groups. We can abstract the circuit structure exhibited by each simplified IR group into something resembling a Tetris block. Then, we choose the ordering based on the circuit-depth cost function of assembling Tetris blocks, with gate cancellation opportunities and qubit routing overhead also taken into account. We primarily have three ingredients:
    \begin{enumerate}
        \item \textit{Endian vectors of circuits and the depth cost function.} As shown in \Cref{fig:empty-end} (a), we define a pair of vectors $e_l$ and $e_r$ for each subcircuit (which would correspond to a simplified IR group in our case). The $i$-th entry of $e_l$ ($e_r$) refers to how many layers one has to traverse starting from the left (right) side before qubit $i$ is acted upon. The layers are defined by grouping neighboring 2Q gates that act on different qubits. See~\Cref{fig:empty-end} (a) for an example of a generic (not necessarily simplified IR group) circuit.
        With this definition, the depth overhead for assembling two subcircuits---the proceeding subcircuit with left-endian vectors $e_r$ and the succeeding subcircuit with left-endian vector $e_l'$ is given by
        \begin{align*}
            \mathrm{cost}_\mathrm{depth} \coloneqq  
                    \left\{
                    \begin{array}{ll}
                    \textsc{sum}(\eRrightPre + \eLeftPost),  \text{ if } \textsc{all}(\eRrightPre[\eLeftPost == 0] > 0) \\\qquad \text{    }\qquad\text{    }\qquad\text{   }\text{and }\textsc{all}(\eLeftPost[\eRrightPre == 0] > 0) \\
                    \textsc{sum}(\eRrightPre + \eLeftPost - 1), \text{ otherwise    }
                    \end{array}
                    \right.
        \end{align*}
        See \Cref{fig:empty-end} (b) for an example. 
        \item \emph{Clifford2Q cancellation.} The simplified IR group usually exposes Hermitian Clifford2Q operators at the two ends of the subcircuit. Therefore some Clifford2Q gate cancellation opportunities could be exploited, which may or may not decrease overall circuit depth, as depicted in \Cref{fig:gate-cancel} (a). We can add this consideration into $\mathrm{cost}_\mathrm{depth}$: if $m$ pairs Clifford2Q cancelled, while (a) without decreasing subcircuit depth, $\mathrm{cost}_\mathrm{depth} \gets \mathrm{cost}_\mathrm{depth} - 2m$; (b) with one side's subcircuit depth decreases, $\mathrm{cost}_\mathrm{depth} \gets \mathrm{cost}_\mathrm{depth} -2m -n$; (c) with both sides' subcircuit depths decrease, $\mathrm{cost}_\mathrm{depth} \gets \mathrm{cost}_\mathrm{depth} - 2m -2n$.
        \item \emph{Consideration of qubit routing overhead.} 
        We try to mitigate the routing overhead for hardware-aware compilation by comparing the qubit interaction graphs of subcircuits. This approach is not limited to specific routing schemes (e.g. three-$\CNOT$ unrolling of $\SWAP$-based routing, ancilla-based bridge gate~\cite{itoko2020optimization}) 
        and hardware topologies. Intuitively, two subcircuits with more \dquote{similar} qubit interaction behaviors require less mapping transition overhead between them, as demonstrated by \Cref{fig:gate-cancel} (b). Therefore, we introduce a factor characterized by the similarity between the qubit interaction graphs of two subcircuits in the cost function:
        \begin{align}
            \mathrm{cost}_\mathrm{depth} \gets \frac 1 s \cdot \mathrm{cost}_\mathrm{depth},\quad s = \sum\nolimits_i \frac{\langle D_i, D_i'\rangle}{\lVert D_i \rVert_2 \lVert D_i' \rVert_2},
        \end{align}
        where $D$ ($D'$) is the distance matrix of the preceding (succeeding) subcircuit's qubit interaction graph of the tail (head) part. The tail (head) is defined as starting from the right (left) of the subcircuit and incorporating more and more 2Q gates until all qubits are acted upon. $D_{i,j}$ represents the shortest path length between qubit $i$ and qubit $j$. $D_i$ is the $i$-th row of $D$.
    \end{enumerate}

    In practice, these IR groups are first pre-arranged in descending order of their weight (subcircuit width). \phoenix\ looks ahead for a certain number of subcircuits to find the one with the minimum $\mathrm{cost}_\mathrm{depth}$ with respect to the last assembled subcircuit. This process iterates until all subcircuits are assembled. 
    


\section{Evaluation}

We evaluate the effectiveness of \phoenix\ across diverse Hamiltonian simulation programs, quantum ISAs, and device topologies. Although \phoenix\ is implemented in Python, it compiles VQA programs of thousands of Pauli strings and more than ten qubits (approximately corresponding to the program size with $10^4$-$10^6$ $\CNOT$ gates in conventional synthesis) in dozens of seconds. Code and data are available on GitHub~\cite{phoenixGitHub}. All experiments are executed on a laptop (Apple M3 Max, 36GB memory).

\subsection{Experimental settings}

    \begin{table}[tbp]
        \centering
        \caption{UCCSD benchmark suite.}
        \setlength{\tabcolsep}{4.2pt}
        \scalebox{0.85}{
            \begin{tabular}{|l|r|r|r|r|r|r|r|}
    \hline
    \textbf{Benchmark} & \textbf{\#Qubit} & \textbf{\#Pauli} & $\mathbf{w_{\max}}$ & \textbf{\#Gate} & \textbf{\#CNOT}  & \textbf{Depth} & \textbf{Depth-2Q} \\
    \hline
    CH2\_cmplt\_BK & 14 & 1488 & 10 & 37780 & 19574 & 23568 & 19399 \\
    \hline
    CH2\_cmplt\_JW & 14 & 1488 & 14 & 34280 & 21072 & 23700 & 19749 \\
    \hline
    CH2\_frz\_BK & 12 & 828 & 10 & 19880 & 10228 & 12559 & 10174 \\
    \hline
    CH2\_frz\_JW & 12 & 828 & 12 & 17658 & 10344 & 11914 & 9706 \\
    \hline
    H2O\_cmplt\_BK & 14 & 1000 & 10 & 25238 & 13108 & 15797 & 12976 \\
    \hline
    H2O\_cmplt\_JW & 14 & 1000 & 14 & 23210 & 14360 & 16264 & 13576 \\
    \hline
    H2O\_frz\_BK & 12 & 640 & 10 & 15624 & 8004 & 9691 & 7934 \\
    \hline
    H2O\_frz\_JW & 12 & 640 & 12 & 13704 & 8064 & 9332 & 7613 \\
    \hline
    LiH\_cmplt\_BK & 12 & 640 & 10 & 16762 & 8680 & 10509 & 8637 \\
    \hline
    LiH\_cmplt\_JW & 12 & 640 & 12 & 13700 & 8064 & 9342 & 7616 \\
    \hline
    LiH\_frz\_BK & 10 & 144 & 9 & 2890 & 1442 & 1868 & 1438 \\
    \hline
    LiH\_frz\_JW & 10 & 144 & 10 & 2850 & 1616 & 1985 & 1576 \\
    \hline
    NH\_cmplt\_BK & 12 & 640 & 10 & 15624 & 8004 & 9691 & 7934 \\
    \hline
    NH\_cmplt\_JW & 12 & 640 & 12 & 13704 & 8064 & 9332 & 7613 \\
    \hline
    NH\_frz\_BK & 10 & 360 & 9 & 8303 & 4178 & 5214 & 4160 \\
    \hline
    NH\_frz\_JW & 10 & 360 & 10 & 7046 & 3896 & 4640 & 3674 \\
    \hline
\end{tabular}    

        }
        \label{tab:uccsd}
        
    \end{table}

    \textbf{Metrics.} We evaluate \phoenix\ using the following metrics: 2Q gate count, 2Q circuit depth, and the algorithmic error for VQA synthesis. Algorithmic error refers to the deviation between the synthesized circuit's unitary matrix and the ideal evolution under the original Hamiltonian, as measured by the infidelity between unitary matrices in our evaluation: $\mathrm{infid} = 1 - \frac{1}{N}|\mathrm{Tr}(U^\dagger V)|$. Notably, we exclude 1Q gates and their count in circuit depth, as 1Q gates are generally considered free resources due to their significantly lower error rates. Additionally, $\CNOT$ is not a native operation on most NISQ platforms, requiring extra 1Q drives before and/or after native 2Q gates (e.g., Cross-Resonance~\cite{rigetti2010fully}, Mølmer-Sørensen~\cite{bruzewicz2019trapped}), making 1Q gate inclusion in metrics potentially misleading.  
    
    \textbf{Baselines.} \tket, \paulihedral, and \tetris are primary baselines to be compared with our method. For \tket, the \code{PauliSimp} and \code{FullPeepholeOptimise} passes are adopted for logical circuit optimization. It is similar to \tket's O3 compilation in which the \code{PauliSimp} pass is particularly effective at optimizing Pauli gadgets. For \paulihedral, the \qiskit\ O2 pass is associated by default because the numerous gate cancellation opportunities exposed by \paulihedral\ necessitate inverse and commutative cancellations. For hardware-aware compilation, all baselines and \phoenix\ are followed by a \qiskit\ O3 pass with SABRE qubit mapping~\cite{li2019tackling}. For QAOA benchmarking, \twoqan~\cite{lao20222qan} is used as the SOTA baseline.

    \textbf{Benchmarks.} We select two representative VQA benchmarks---(1) \emph{UCCSD}: A set of UCCSD ansatzes, including \CHtwo, \HtwoO, \LiH, and \NH: 4 categories of molecule simulation programs. Each category is generated with STO-3G orbitals~\cite{hehre1969self}, encoded by Jordan-Wigner (JW)~\cite{jordan1928paulische} and Bravyi-Kitaev (BK)~\cite{bravyi2002fermionic} transformations, in turn approximated by complete or frozen-core orbitals. Details are shown in \Cref{tab:uccsd}. (2) \emph{QAOA}: A set of 2-local Hamiltonian simulation programs corresponding to random graphs and regular graphs, for which the description and evaluation results are shown in \Cref{tab:qaoa}.

\subsection{Main results}

    \begin{figure}[tbp]
        \centering
        \includegraphics[width=\columnwidth,trim={0 0.3cm 0 0.2cm},clip]{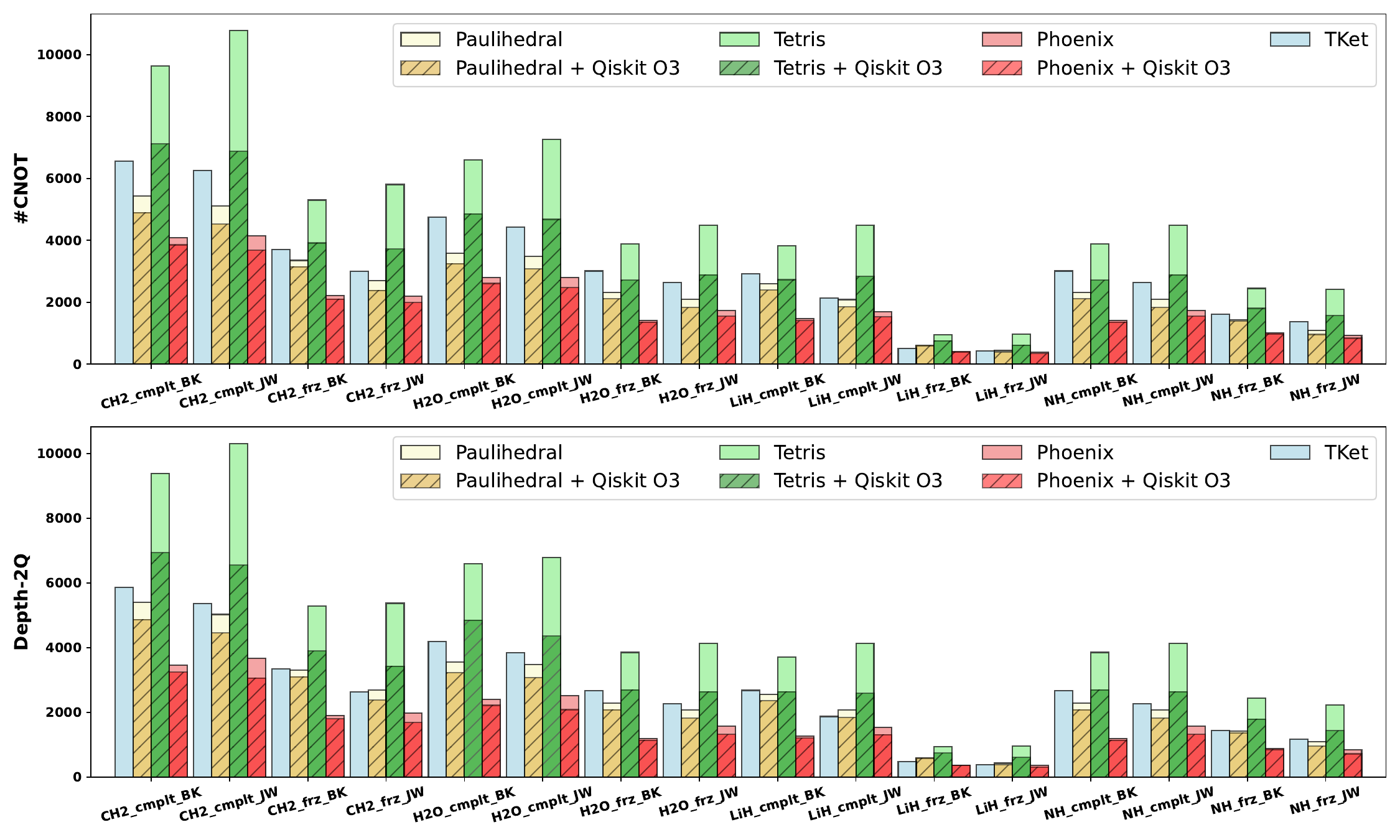}
        \caption{Logical-level compilation (all-to-all topology).}
        \label{fig:all2all}
    \end{figure}

    \begin{table}[tbp]
        \centering
        \caption{Average (Geometric-mean) optimization rates on UCCSD.}
        \scalebox{0.95}{
            \begin{tabular}{|l|c|c|}
    \hline
    \textbf{Compiler}          & \textbf{\#CNOT opt.}                        & \textbf{Depth-2Q opt.}                     \\ 
    \hline
    \tket                       & 33.07\%                                    & 30.14\%                                   \\ 
    \hline
    \paulihedral                & 28.41\%                                    & 29.07\%                                   \\ 
    \hline
    \multirow{2}{*}{\paulihedral\ + O3}    
                               & 25.72\%                                     & 26.3\%                                    \\ 
                               & (-8.54\% v.s. no O3)                 & (-8.6\% v.s. no O3)               \\ 
    \hline
    \tetris                     & 53.66\%                                    & 53.26\%                                   \\ 
    \hline
    \multirow{2}{*}{\tetris\ + O3}         
                               & 36.73\%                                     & 36.37\%                                   \\ 
                               & (-30.94\% v.s. no O3)                & (-31.08\% v.s. no O3)              \\ 
    \hline
    \phoenix                    & \textbf{21.13\%}                                    & \textbf{19.3\%}                                   \\ 
    \hline
    \multirow{2}{*}{\phoenix\ + O3}        
                               & \textbf{19.53\%}                                     & \textbf{17.3\%}                                   \\ 
                               & (\textbf{-6.6\%} v.s. no O3)                 & (\textbf{-8.44\%} v.s. no O3)               \\ 
    \hline
\end{tabular}
            
        }
        \label{tab:uccsd-avg}        
    \end{table}

    \Cref{fig:all2all} and \Cref{tab:uccsd-avg} illustrates the main benchmarking results regarding logical-level compilation:
    \begin{enumerate}
        \item \phoenix\ significantly outperforms baselines across all benchmarks, with an average (geometric-mean) 21.12\% and 19.29\% optimization rate in \#$ \CNOT $ and Depth-2Q, respectively, relative to original circuits.\footnote{For example, the \#$ \CNOT $ optimization rate is defined as $\frac{\#\CNOT_\textrm{after}}{\#\CNOT_\textrm{before}}$.} That is mostly attributed to the group-wise BSF simplification mechanism, as \phoenix\ adopts the same IR grouping method as \paulihedral\ and \tetris.
        \item \tetris\ performs the worst, falling far behind \tket, \paulihedral, and \phoenix. This is because \tetris\ focuses primarily on co-optimization techniques to reduce $\SWAP$ gates during qubit routing, rather than logical-level synthesis.
        \item We also compare \paulihedral/\tetris/\phoenix\ with and without \qiskit\ O3, to evaluate their high-level optimization capabilities. The improvement in using \qiskit\ O3 for \paulihedral\ and \tetris\ is more pronounced than for \phoenix. Therefore, \phoenix's high-level optimization strategy is more impressive, leaving less optimization space for \qiskit\ O3.
    \end{enumerate}

\subsection{Hardware-aware compilation}

    \begin{figure}[tbp]
        \centering
        \includegraphics[width=\columnwidth,trim={0 0.3cm 0 0.2cm},clip]{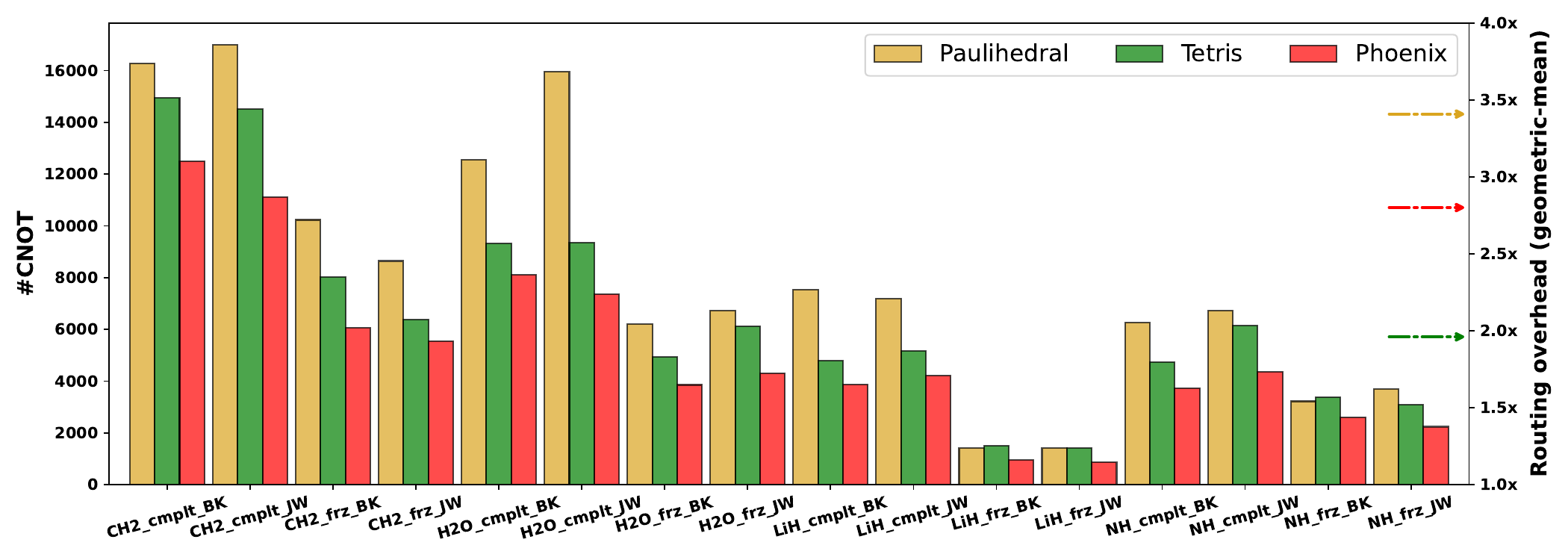}
        \caption{Hardware-aware compilation on heavy-hex topology. Three dashed lines represent the average multiples of \#$\CNOT$ within circuits after mapping relative to those after logical optimization, for \paulihedral\ (gold), \tetris\ (green), and \phoenix\ (coral), respectively.}
        \label{fig:manhattan}
    \end{figure}

    We use the heavy-hex topology, specifically a 64-qubit coupling graph of IBM's Manhattan processor~\cite{mooney2021whole}, for hardware-aware compilation. Results are shown in \Cref{fig:manhattan}, where \tket\ is excluded due to its significantly worse performance compared to others. Despite focusing primarily on high-level logical program optimization with humble hardware-aware co-optimization, \phoenix\ still outperforms baselines, reducing $\CNOT$ gate count by 35.26\% (21.69\%) and 2Q circuit depth by 43.54\% (27.68\%) on average, compared to \paulihedral\ (\tetris). Considering the qubit mapping transition overhead mitigation within its IR group ordering process, \phoenix\ induces 2.83x \#$ \CNOT $ on average after hardware mapping. It is better than \paulihedral\ while worse than \tetris, as \tetris\ specializes in $\CNOT$ cancellation for $ \SWAP $-based routing. Consequently, even on limited-topology devices, \phoenix\ effectively manages routing overhead and surpasses co-design local optimization strategies.

\subsection{Comparison in diverse ISAs}

    \begin{table*}[tbp]
        \centering
        \caption{Comparison for diverse ISAs with all-to-all and limited topology.}
        \scalebox{0.9}{
            \begin{tabular}{|l|c|c|c|c|c|c|c|c|}
    \hline
     & \multicolumn{2}{|c|}{\textbf{CNOT ISA (all-to-all)}} & \multicolumn{2}{|c|}{\textbf{SU(4) ISA (all-to-all)}} & \multicolumn{2}{|c|}{\textbf{CNOT ISA (heavy-hex)}} & \multicolumn{2}{|c|}{\textbf{SU(4) ISA (heavy-hex)}} \\ \hline
    \textbf{\phoenix's opt. rate} & {\#CNOT} & {Depth-2Q} & {\#SU(4)} & {Depth-2Q} & {\#CNOT} & {Depth-2Q} & {\#SU(4)} & {Depth-2Q} \\ \hline
    \phoenix\ v.s. \tket          & 63.88\%       & 64.05\%           & \textbf{56.06\%} & \textbf{54.25\%} & \textbf{41.05\%} & \textbf{48.48\%} & 44.65\%       & 50.85\%           \\ \hline
    \phoenix\ v.s. \paulihedral   & 82.14\%       & 73.4\%           & \textbf{75.6\%} & \textbf{65.24\%} & 63.03\%         & 54.7\%         & \textbf{40.16\%} & \textbf{35.16\%} \\ \hline
    \phoenix\ v.s. \tetris        & 57.53\%       & 53.08\%           & \textbf{56.57\%} & \textbf{50.58\%} & 76.76\%         & 71.43\%         & \textbf{62.74\%} & \textbf{58.9\%} \\ \hline
\end{tabular}

        }
        \label{tab:isa}
    \end{table*}

    Quantum ISA, or the native gate set in a narrow sense, serves as an interface between software and hardware implementation. For a specific quantum ISA, the adopted 2Q gate dominates the accuracy and difficulty of hardware implementation, as well as the theoretical circuit synthesis capabilities. While traditional ISAs typically consist of 1Q gates and $\CNOT$-equivalent 2Q gates, recently some works propose integrating complex and even continuous 2Q gates into the ISA design, such as the $\mathrm{XY}$ gate family~\cite{abrams2020implementation}, the fractional/partial $\mathrm{ZZ}$ and $\mathrm{MS}$ gates~\cite{ibmFractionalGates,ionqPartialGates}, and the AshN gate scheme which considers all possible 2Q gates within the $\SUfour$ group as the ISA~\cite{chen2024one}.    
    Therefore, we further compare \phoenix\ with baselines in different ISAs to showcase its ISA-independent compilation advantage.
    
    The ISA-independent advantage is best shown by choosing the most expressive $\SUfour$ ISA. We evaluate it with both all-to-all and heavy-hex topologies, as the same as the evaluation above for $\CNOT$ ISA. For logical-level compilation, \phoenix\ directly generates $ \SUfour $-based circuits via its BSF simplification algorithm, while baselines (\paulihedral\ and \tetris\ equipped with \qiskit\ O3 by default) require an additional \dquote{rebase} (or \dquote{transpile}) step to convert $ \CNOT $-based circuits to $ \SUfour $-based ones. For hardware-aware compilation, all compilers include a rebase step, following the \qiskit\ O3 hardware-aware compilation pass. Detailed outcomes are summarized in \Cref{tab:isa}, highlighting the geometric-mean relative optimization rates of \phoenix\ compared to the baselines' results.

    Again, \phoenix\ significantly outperforms baselines when targeting $ \SUfour $ ISA. The optimization rates relative to baselines are more impressive than those in $ \CNOT $ ISA, despite the baselines incorporating sophisticated optimization techniques specifically designed for the $ \CNOT $ ISA. For instance, the multiple of \phoenix's \#2Q relative to \paulihedral's in $ \CNOT $ ISA is 82.12\% (62.38\%), whereas this value decreases to 75.57\% (39.84\%) in $ \SUfour $ ISA for hardware-agnostic (hardware-aware) compilation. One exception is the hardware-aware compilation comparison with \tket, as \tket's hardware-aware compilation in the $ \CNOT $ ISA generates circuits with much larger 2Q gate count and circuit depth than other compilers, and there are numerous 2Q subcircuit fusing opportunities such that the rebased circuits involves much fewer $ \SUfour $ gates.





\subsection{QAOA benchmarking}

    \begin{figure}[tbp]
        \centering
        \includegraphics[width=\columnwidth,trim={0 0.3cm 0 0.2cm},clip]{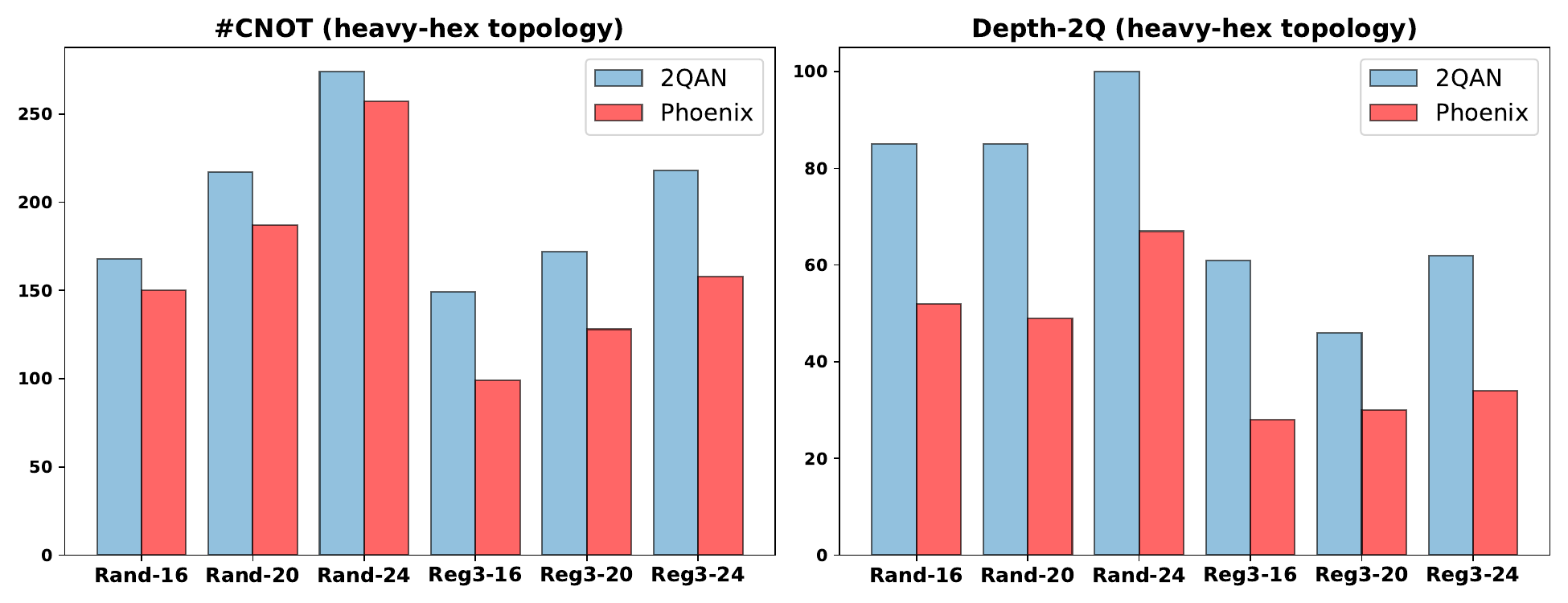}
        \caption{QAOA benchmarking}
        \label{fig:qaoa}
    \end{figure}

    \begin{table}[btp]
        \centering
        \caption{QAOA benchmarking versus 2QAN.}
        \setlength{\tabcolsep}{3.8pt}
        \scalebox{0.76}{
            \begin{tabular}{|l|r|r|r|r|r|r|r|r|r|r|}
    \hline
    \multicolumn{2}{|c|}{\textbf{QAOA}} & \multicolumn{2}{c|}{\textbf{\#CNOT}} & \multicolumn{2}{c|}{\textbf{Depth-2Q}} & \multicolumn{2}{c|}{\textbf{\#SWAP}} & \multicolumn{2}{c|}{\textbf{Routing overhead}} \\ 
    \hline
    Bench. &  \#Pauli & 2QAN & Phoenix & 2QAN & Phoenix & 2QAN & Phoenix & 2QAN & Phoenix \\
    \hline
    Rand-16 & 32 & 168 & 150 & 85 & 52 & 37 & 29 & 2.62x & 2.34x \\
    \hline
    Rand-20 &  40 & 217 & 187 & 85 & 49 & 47 & 39 & 2.71x & 2.34x \\
    \hline
    Rand-24 &  48 & 274 & 257 & 100 & 67 & 63 & 56 & 2.85x & 2.68x \\
    \hline
    Reg3-16 &  24 & 149 & 99 & 61 & 28 & 44 & 17 & 3.10x & 2.06x \\
    \hline
    Reg3-20 &  30 & 172 & 128 & 46 & 30 & 46 & 23 & 2.87x & 2.13x \\
    \hline
    Reg3-24 &  36 & 218 & 158 & 62 & 34 & 62 & 30 & 3.03x & 2.19x \\
    \hline
    \multicolumn{2}{|c|}{\emph{Avg. improv.}} & \multicolumn{2}{c|}{-16.7\%} & \multicolumn{2}{c|}{-40.8\%} & \multicolumn{2}{c|}{-29.41\%} & \multicolumn{2}{c|}{-16.59\%} \\
    \hline
\end{tabular}
    
        }
        \label{tab:qaoa}
    \end{table}

    For QAOA benchmarking, we focus on the performance in hardware-aware compilation, since the 2Q gate count cannot be reduced and minimizing circuit depth is easy in logical-level compilation. Both \twoqan\ and \phoenix\ can generate depth-optimal QAOA circuits at the logical level in our field test. \Cref{fig:qaoa} and \Cref{tab:qaoa} illustrate compilation results on the heavy-hex topology, across six QAOA programs corresponding to both random graphs (each node with degree 4) and regular (each node with degree 3) graphs, with qubit sizes of 16, 20, and 24. \phoenix\ outperforms \twoqan\ across all benchmarks in all metrics (such as \#$ \CNOT $, \#$ \SWAP $), especially in Depth-2Q, with an average 40.8\% reduction compared to \twoqan. These results further demonstrate the effectiveness of the routing-aware IR group ordering method in \phoenix.

\subsection{Algorithmic error analysis}

    \begin{figure}[tbp]
        \centering
        \includegraphics[width=\columnwidth,trim={0 0.3cm 0 0.2cm},clip]{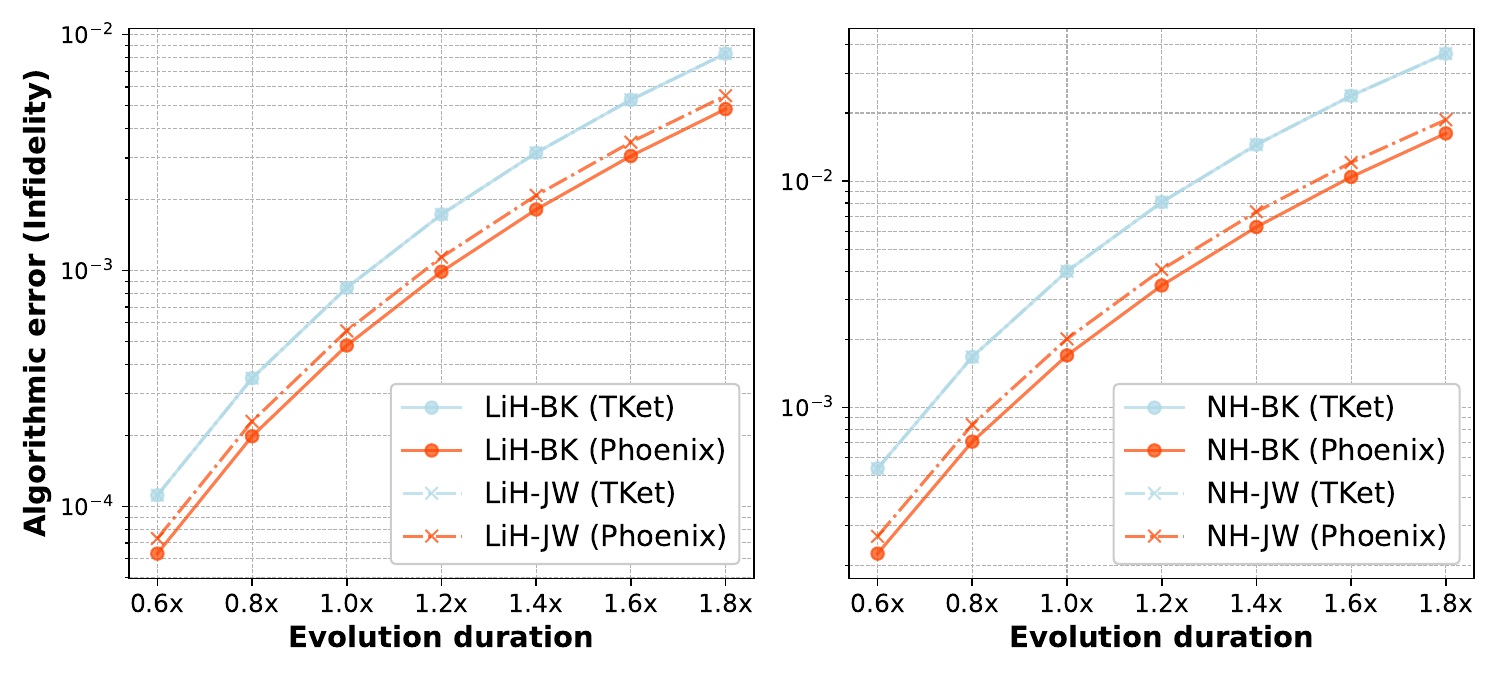}
        \caption{Algorithmic error comparison of \LiH\ and \NH\ simulation.  }
        \label{fig:algo-err}
    \end{figure}

    We further highlight \phoenix's advantage in reducing the algorithmic error. We select UCCSD benchmarks with qubits no more than 10 for evaluation, within the matrix computation capabilities of standard PCs. We rescale the coefficients of Pauli strings to control their algorithmic errors within $5\times 10^{-5}$ to $10^{-2}$, which corresponds to different evolution durations in molecular simulation, as suggested in \Cref{fig:algo-err}. In contrast to \tket, \phoenix\ typically leads to lower algorithmic errors for both JW and BK encoding schemes. Although this improvement is program-specific, it is more significant for the Pauli string patterns of BK than those of JW, with 57\% (42.7\%) and 49.5\% (34.1\%) for \NH\ (\LiH) simulation, respectively. As they adopt the same Pauli string blocking approach, we expect the algorithmic errors resulting from \paulihedral\ and \tetris\ to be comparable to \phoenix, and so are not shown in \Cref{fig:algo-err}. As a result, the impressive algorithmic error reduction effect of \phoenix\ brings us closer to a possible quantum advantage on computational chemistry problems.




\section{Conclusion and Outlook}
Mainstream circuit synthesis approaches rely on pattern rewrite rules, often restricted to small-scale, local optimizations. In contrast, we present \phoenix, a framework leveraging high-level Pauli-based IR for Hamiltonian simulation, one of the most prominent NISQ applications. \phoenix\ outperforms all SOTA VQA compilers across diverse programs and hardware platforms, showcasing its unmatched performance and versatility. This work not only bridges the gap between impactful quantum applications and physically implementable solutions but also prompts a re-evaluation of compiler optimization.


Beyond producing significantly optimized circuits, high-level IRs, as a refined abstraction layer on top of the traditional quantum ISA, create more opportunities for efficient synthesis and effective hardware mapping. Furthermore, they can act as intermediate building blocks, guiding the design of quantum processors and corresponding control schemes---potentially incorporating multi-qubit control---to implement these high-level IRs more efficiently.

\section*{Acknowledgement}
    This work is supported by AI Chip Center for Emerging Smart Systems (ACCESS), sponsored by InnoHK funding, Research Grants Council of HKSAR (No. 16213824) and National Key Research and Development Program of China (No. 2023YFA1009403), National Natural Science Foundation of China (No. 12347104), Beijing Natural Science Foundation (No. Z220002).
    DD would like to thank God for all of His provisions.


\bibliographystyle{IEEEtran}
\bibliography{reference}



\end{document}